\documentclass[lettersize,journal]{IEEEtran}
\usepackage{cite}
\usepackage{amsmath,amssymb,amsfonts}
\usepackage{graphicx}
\usepackage{textcomp}
\usepackage[table]{xcolor}
\usepackage{tcolorbox}   
\usepackage{color}
\usepackage{framed}
\usepackage{subcaption}
\definecolor{shadecolor}{rgb}{0.92,0.92,0.92}
\usepackage{diagbox}
\usepackage[noend]{algpseudocode}
\usepackage{algorithmicx,algorithm}
\usepackage{tabularx}
\usepackage{colortbl}

\usepackage{titlesec}

\titlespacing{\subsection}{0pt}{\parskip}{0.5\parskip}

\newtheorem{theorem}{Lemma}
 
\newcommand{\llbracket}{[\![}
\newcommand{\rrbracket}{]\!]}
\newcommand{\bc}[1]{\mbox{\boldmath $\mathcal{#1}$}}
\newcommand{\mf}[1]{\mathbf{#1}}
\newcommand{\mb}[1]{\mathbb{#1}}

\newcommand{\HH}{\mathrm{H}}
\newcommand{\T}{\mathrm{T}}

\begin{document}
\title{Estimating Channels With Hundreds of Sub-Paths for  MU-MIMO Uplink: \\ A Structured High-Rank Tensor Approach }


\author{Panqi Chen and Lei Cheng
	\thanks{P. Chen and L. Cheng are with the College of Information Science and Electronic Engineering, Zhejiang University, Hangzhou, China. E-mails: \{Panqi$\_$Chen, lei$\_$cheng\}@zju.edu.cn }
	}



\maketitle

\begin{abstract}
	This letter introduces a structured high-rank tensor approach for estimating sub-6G uplink channels in  multi-user  multiple-input and multiple-output (MU-MIMO) systems. To tackle the difficulty of channel estimation in sub-6G bands with hundreds of sub-paths, our approach fully exploits the physical structure of channel and establishes the link between sub-6G channel model and a high-rank four-dimensional (4D) tensor Canonical Polyadic Decomposition (CPD) with three factor matrices being Vandermonde-constrained. Accordingly, a stronger uniqueness property is derived in this work.	
This model
supports an efficient one-pass algorithm for estimating sub-path
parameters, which ensures plug-in compatibility with the widely-used baseline.  Our method performs much better than the state-of-the-art tensor-based techniques on the simulations adhering to the 3GPP 5G protocols.
\end{abstract}

\begin{IEEEkeywords}
	 MU-MIMO, uplink channel estimation,  structured high-rank tensor decomposition. 
\end{IEEEkeywords}

\section{Introduction}

\IEEEPARstart{M}ulti-user multiple-input and multiple-output (MU-MIMO) communications holds the promise of significantly boosting uplink capacity to meet the growing demands of 5.5G/6G applications, including  broadband Internet of Things \cite{ucbc2}. Despite its enormous potential,  MU-MIMO communications presents significant challenges to the physical-layer signal processing at the base station, especially in estimating sub-6G channels for multiple users simultaneously. 

Sub-6G bands, with  superior coverage and lower susceptibility to interference than higher frequency bands, are  primarily  used in real-life wireless communication systems\cite{sub6g}. However, channel estimation in sub-6G bands is challenging due to the typically hundreds of sub-paths involved\cite{sub6g}. In  MU-MIMO systems, where multiple users in various cells transmit data simultaneously, interference of adjacent cells can further complicate channel estimation.

To address these challenges, we propose a structured high-rank tensor approach to estimate sub-6G uplink channels in  MU-MIMO systems. We focus on a MIMO-OFDM system with a polarized planar antenna array, which is widely used in real-life wireless systems\cite{sub6g}.  Previous works have shown that MIMO-OFDM channel models can be recast as tensor canonical polyadic decompositions (CPDs)\cite{Ftcom,rztcom,Ytcom,DD,ISC,MIMO}, leveraging the associated uniqueness properties to facilitate parameter estimation. However, these studies primarily consider mmWave channels with only a few sub-paths. While recent works \cite{ISC, DD} have explored the Vandermonde structure of a {\it single factor matrix} in the channel CPD\cite{tensor} model to enable stronger uniqueness properties, their findings face challenges in solving our problem involving hundreds of sub-paths.

To better utilize the algebraic structure of channel models, we propose to model the channel as a fourth-order CPD with {\it three factor matrices} being Vandermonde-constrained. This structure significantly enhances the uniqueness of the channel tensor model, which is theoretically and empirically verified in this letter, thus making identifying hundreds of sub-paths viable. To exploit the Vandermonde structure with low-cost computation, we extend the tensor ESPRIT method\cite{vand}, which is a non-iterative algorithm originally designed for a third-order tensor CPD with one Vandermonde factors, to acquire the channel parameters from a fourth-order tensor CPD model with three Vandermonde factors. Note that unlike recent related works \cite{ISC,DD} that focus on single-user downlink transmissions with fewer sub-paths, our letter addresses challenging MU-MIMO uplink transmissions with {\it numerous sub-paths} and under {\it notorious interferences}.

Simulations that strictly follow the 3GPP protocols\cite{3GPP38901,3GPP38214,3GPP38211}, including the pilot signal (DeModulation Reference Signal, DMRS), the channel model (3GPP  Cluster Delay Line (CDL) model), and the baseline pipeline are conducted to evaluate the proposed algorithm. Numerical results demonstrate that our proposed method can significantly reduce the channel estimation error, making it well-suited for real-life  applications.

{\it Notations:} Lower-case, upper-case bold  and upper-case bold calligraphic letters (e.g. $\mf x$, $\mf X$ and $\bc{X}$) are used to denote vectors, matrices, higher-order tensors. Operations $\odot$, $\ast$,  $\otimes$ denote  Khatri-Rao, Hadmard  and Kronecker product respectively.

\section{System Model and Problem Formulation}
\label{sec2}

A MU-MIMO-OFDM system is considered, where the base station (BS) at the targeted cell has a multi-polarized uniform planar array (UPA) with $N_{\text{row}} \times N_{\text{col}}$ antennas and serves $U$ user equipments (UEs). Each UE is assumed to have a single antenna. The frequency-domain received data model at the $t$-th  symbol duration can be expressed as:
\vspace{-2mm}
\begin{equation}
	\label{eq1}
	\mf Y(t) =\sum_{u=1}^{U} \mf H^{u}(t) \mf S^{u}(t) + \mf N(t) + \mf I (t), 
\end{equation}
where $\mf H^{u}(t) \in \mb C^{N_{\text{col}}N_{\text{row}} N_{\text{pol}} \times K}$ denotes the channel matrix of the $u$-th UE at the $t$-th symbol duration; $ \mf S^{u}(t)=\text{diag}\{s^{u}_1(t), s^{u}_2(t),  ..., s^{u}_K(t)\} \in \mb C^{K \times K}$ is the transmitted symbol matrix of the $u$-th UE, with $s^{u}_k(t)$ being  transmitted symbol(e.g., QPSK, QAM) at the $k$-th subcarrier and  the $t$-th duration;  $\mf N(t) \in \mb C^{N_{\text{col}}N_{\text{row}}N_{\text{pol}} \times K}$ denotes the additive white Gaussian noise (AWGN) and $\mf I(t) \in \mb C^{N_{\text{col}}N_{\text{row}}N_{\text{pol}} \times K}$ denotes the interferences from other cells.

In this study, we adopt the CDL channel model specified in the 3GPP protocol \cite{3GPP38901}. Specifically, the frequency-domain uplink channel coefficient of $u$-th UE is modeled as:
\vspace{-2mm}
\begin{align}
	\label{eq6}
	&h^u_{ n_{\text{pol}}, n_{\text{row}}, n_{\text{col}},  k} (t)  =\sum_{l=1}^{L}  
	\alpha^{u}_{l,n_{\text{pol}}} e^{-2j\pi \tau^{u}_l(k\Delta f) } \nonumber \\
	&~~~~~~~~~~~~~~\times e^{ \frac{2j\pi f_{\text{c}} n_{\text{col}} d_{\text{col}}}{\text{c}} \sin\phi_{\text{AOA},l}^{u} }    e^{ \frac{2j\pi f_{\text{c}} n_{\text{row}} d_{\text{row}}}{\text{c}}\cos\theta_{\text{ZOA},l} ^{u} },
\end{align}
where $ (n_{\text{col}},n_{\text{row}}),k, n_{\text{pol}}$ denote  BS antenna index, subcarrier index, and direction of polarization, respectively.  $\Delta f$ is subcarrier interval; $f_{\text{c}}$ is the carrier frequency; and  $\text{c}$ is the speed of electromagnetic wave. $d_{\text{col}}$ and $d_{\text{row}}$ denote column space and row space between antennas, respectively.  The channel coefficient $h^u_{ n_{\text{pol}}, n_{\text{row}}, n_{\text{col}},  k} (t)$ can be interpreted as the summation of $L$  sub-path's responses. For the $l$-th sub-path:
 $ \alpha^{u}_{l,n_{\text{pol}}}$ and $\tau^u_l$ represent  sub-path gain and time delay from the $u$-th UE to all antennas in BS, respectively; 
 $\phi_{\text{AOA},l}^{u}$ and $\theta_{\text{ZOA},l} ^{u}$  are Azimuth angle of  arrival (AOA) and Zenith angles of arrival (ZOA), respectively.

Our objective is to estimate the  channel coefficients for all users, denoted by $\{ \mf {H}^{u}\}_{u=1}^{U}$,  using the $T_p$ symbols of pilot signals (i.e., $\{\mf S^{u}_{\text{pilot}}(t)\}_{t=1}^{T_p}$) and the received data  (i.e., $\{\mf Y(t)\}_{t=1}^{T_p}$) at BS. The channels are assumed to be invariant  over the duration of $T_p$ pilot signals.   This leads us to formulate the following optimization problem:
\vspace{-2mm}
\begin{equation}
	\label{eq7}
	\min_{ \{ \mf {H}^{u}\}_{u=1}^{U} } \sum_{t=1}^{T_p} \left \| \mf Y(t) -\sum_{u=1}^{U} \mf H^{u} \mf S^{u}_{\text{pilot}}(t) \right \|_{\text{F}}^2 . 
\end{equation}
However, the optimization problem (\ref{eq7}) poses two significant challenges. First, the channel matrices to be estimated have a large number of parameters, i.e., $U N_{\text{col}}N_{\text{row}} N_{\text{pol}} K$ (up to millions in typical practical systems\cite{3GPP38901,3GPP38214,3GPP38211}), while the received data only provide few entries,  resulting in an ill-conditioned and under-determined problem. Second, the channel matrices of different UEs $\{ \mf {H}^{u}\}_{u=1}^{U}$ are coupled in problem (\ref{eq7}), making it challenging to uniquely identify them.

\section{Baseline and The Proposed Method }
\label{se3}

\subsection{Baseline}
The baseline approach to solve problem (\ref{eq7}) involves the  least-squares estimation method and the linear interpolation technique, while utilizing the well-designed pilot signals known as  DMRS \cite{3GPP38211}. These  pilot signals are carefully designed to ensure optimal utilization of the time-frequency resource, guaranteeing orthogonality across the data streams from the $U$ UEs\footnote{{ More details can be referred to Table 6.4.1.1.3-2 of \cite{3GPP38211}}.}. Specifically, for each UE,  a specific set of indices (referred to as a comb) from the $K$ subcarriers of the data stream are firstly acquired. The comb indicates the subcarriers on which the UE transmits its pilot data. A least-squares estimation algorithm is then applied, resulting in  incomplete channel estimations denoted as $\{\mf {\hat{H}}^u_{\text{comb}} \in \mb{C}^{  N_{\text{row}}N_{\text{col}} N_{\text{pol}} \times N_{\text{sc,eff}}}   \}_{u=1}^{U}$. Here, $N_{\text{sc,eff}}$  represents the effective number of subcarriers within the comb. Ultimately,  full channel estimation $\{\mf {\hat{H}}^u \in \mb{C}^{  N_{\text{row}}N_{\text{col}} N_{\text{pol}} \times K }\}_{u=1}^{U}$ is achieved by linearly interpolating the missing values.

\subsection{ The Proposed Method}
The baseline method is highly susceptible to noise, resulting in a significant degradation of performance in lower SNR regimes. This drawback stems from two factors: First, the baseline method overlooks the inherent multi-dimensional (M-D) structure of the channel (time, space, and frequency); Second, the trivial linear interpolation used  tends to excessively fit the noise, exacerbating the issue.

To unveil the M-D structure of channels, recent works\cite{Ftcom,rztcom,Ytcom,MIMO} have treated received signals as  third-order  CP tensors  but they do not exploit its Vandermonde structure. Ref.\cite{DD}  utilized the compressed tensor decomposition (CTD) method to uniquely estimate channels,  leveraging one Vandermonde factor matrix. However, the uniqueness properties of these previous approaches are not sufficiently strong to handle sub-6G channels. In our study, we model the polarization as a factor matrix and discovered that it leads to a  fourth-order tensor CPD \textit{with three factor matrices being Vandermonde constraint}, resulting a {\it much stronger uniqueness property} (see Lemma 1 presented in the next page). 
Specifically,  we propose to reshape each channel matrix  $\mf H^{u} \in \mb{C}^{  N_{\text{row}}N_{\text{col}}N_{\text{pol}} \times K }$ to  $ \bc{H}^u \in \mb{C}^{N_{\text{col}} \times N_{\text{row}} \times K \times N_{\text{pol}}}$, and it can be shown that $\bc{H}^u $ can be expressed as follows:
\vspace{-2mm}
\begin{equation}
	\label{eq8}
	\bc{H}^u=\llbracket \mf{A}^u(\phi),\mf{A}^{u}(\theta),\mf{D}^{u}, \mf{P}^{u}  \rrbracket,  \bc{H}^u \in \mb{C}^{N_{\text{col}} \times N_{\text{row}} \times K \times N_{\text{pol}}},
\end{equation}
where $\mf{A}^u(\phi),\mf{A}^{u}(\theta),\mf{D}^{u},\mf{P}^{u} $ are factor matrices being:
\begin{align}
	\label{model}
		&[\mf{A}^u(\phi)]_{n_{\text{col}},l}=e^{ \frac{2j\pi f_{\text{c}} (n_{\text{col} }-1) d_{\text{col}}}{\text{c}} \sin\phi^u_{\text{AOA},l} } , \mf{A}^u(\phi) \in\mb{C}^{N_{\text{col}}\times L}, \nonumber \\
		&[\mf{A}^{u}(\theta)]_{n_{\text{row}},l}= e^{ \frac{2j\pi f_{\text{c}} (n_{\text{row}}-1) d_{\text{row} }}{\text{c}}\cos\theta_{\text{ZOA},l} ^{u}} ,\mf{A}^u(\theta)  \in \mathbb{C}^{N_{\text {row }} \times L}, \nonumber \\
		&[\mathbf{D}^{u}]_{k, l}=e^{-j 2 \pi \tau^u_l(k-1)},  \mathbf{D}^{u} \in \mathbb{C}^{K \times L}, \nonumber \\
		&[\mf{P}^{u}]_{n_{\text{pol}}, l}= \alpha^u_{l,n_{\text{pol}}}, \mathbf{P}^{u} \in \mathbb{C}^{N_{\text {pol}} \times L}.  
\end{align}
From (\ref{model}), we can observe that all the factor matrices are determined by $\{ \theta_{\text{ZOA},l} ^{u}, \phi^u_{\text{AOA},l}, \alpha^u_{l,n_{\text{pol}}}, \tau^u_l \}_{l=1}^L$. If we can estimate these parameters, the channel tensor can be reconstructed accordingly.
Using the fourth-order tensor channel  model (4),  we  reformulate problem (\ref{eq7}) in tensor form as follows:
\vspace{-2mm} 
\begin{align}
	\label{eq12}
	&\min_{ \{   \{ \theta_{\text{ZOA},l} ^{u}, \phi^u_{\text{AOA},l}, \alpha^u_{l,n_{\text{pol}}}, \tau^u_l \}_{l=1}^L   \}_{u=1}^U   } \nonumber \\
	&\sum_{t=1}^{T_p} \left \| \bc Y(t) -\sum_{u=1}^{U}  \llbracket \mf{A}^u(\phi),\mf{A}^{u}(\theta),\mf S^{u}_{\text{pilot}}(t)\mf{D}^{u}, \mf{P}^{u}  \rrbracket  \right \|_{\text{F}}^2.
\end{align}
Such a re-parameterization significantly reduces the number of unknowns, e.g., from 1,000,000 to 30,000 under typical settings of real-world systems (see Table~\ref{t1}). This reduction addresses the first challenge in solving problem (3). Additionally, the use of a 4D tensor structure for the channel holds promise in mitigating the issue of noise overfitting encountered by the baseline approach. 

Since  baseline approach has tackled the coupling of different UE's channels utilizing orthogonal time-frequency pilot resources, we propose to leverage the baseline approach  and subsequently employ the tensor method for refinement. In this way, our approach can be directly plugged into the pipeline of the existing baseline, showing its great compatibility.  We first reshape the incomplete channel estimations   $\{\mf {\hat{H}}^{u}_{\text{comb}}\}_{u=1}^{U}$ to fourth-order tensors $\{\bc {\hat{H}}^{u}_{\text{comb}}\}_{u=1}^{U} \in \mb{C}^{N_{\text{col}} \times N_{\text{row}} \times N_{\text{sc,eff}} \times N_{\text{pol}}}$ and then solve the following optimization problem:
\vspace{-3mm} 
\begin{align}
	\label{eq15}
	&\min_{  \{   \{ \theta_{\text{ZOA},l} ^{u}, \phi^u_{\text{AOA},l}, \alpha^u_{l,n_{\text{pol}}}, \tau^u_l \}_{l=1}^L   \}_{u=1}^U    } \nonumber \\
	&\sum_{u=1}^U  \left \| \bc{\hat{H}}_{\text{comb}}^u-	\llbracket  \mf{A}^u(\phi),\mf{A}^{u}(\theta),\mf{D}_{\text{comb}}^{u},  \mf{P}^{u} \rrbracket \right \|_{\text{F}}^2,
\end{align}
where  $\mf{D}_{\text{comb}}^{u} =\mf{D}^{u}(:,\mf c) $ with $\mf c$ representing the index set of comb subcarriers,  including $N_{\text{sc,eff}}$ items. Note that $U$ users have been separated leveraging the DMRS and thus solving problem (\ref{eq15}) is equivalent to solving $U$  subproblems separately:
\vspace{-7mm} 
\begin{align}
	\label{eq15a}
	&\min_{     \{ \theta_{\text{ZOA},l} ^{u}, \phi^u_{\text{AOA},l}, \alpha^u_{l,n_{\text{pol}}}, \tau^u_l \}_{l=1}^L      } \nonumber \\
	& \left \| \bc{\hat{H}}_{\text{comb}}^u-	\llbracket  \mf{A}^u(\phi),\mf{A}^{u}(\theta),\mf{D}_{\text{comb}}^{u},  \mf{P}^{u} \rrbracket \right \|_{\text{F}}^2,
\end{align}
At first glance, problem~\eqref{eq15a} appears to be a typical tensor-based multidimensional harmonic retrieval problem, and many existing tensor decomposition algorithms can be applied\cite{ALS}.  Among these methods, the alternating least squares (ALS) algorithm\cite{ALS} is the most commonly employed\cite{Ftcom,rztcom,Ytcom}, which estimates one factor matrix while keeping the others fixed. 
However, directly apply ALS-based methods to our sub-6G  MU-MIMO setting is not a good option.  
The reason is that previous works  primarily focused on mmWave channel estimation, where the channel is assumed to have a small number of sub-paths (e.g., 5-10). Thus, the associated low-rank tensor CPD problem is well-conditioned so that ALS can estimate factor matrices with a uniqueness guarantee. Moreover, Refs.\cite{DD, ISC} adopted the smoothed ESPRIT algorithm to enhance the uniqueness property with one factor matrix being Vandermonde constrained.
However, in our applications, the channel typically consists of hundreds of sub-paths, leading to a high-rank tensor CPD. Without fully exploiting the structure of tensor decomposition, the uniqueness of factor matrices cannot be guaranteed. Consequently, accurately retrieving these parameters becomes impossible.

Upon closer inspection of problem~\eqref{eq15a}, we observe that it differs from existing tensor-based channel estimation\cite{Ftcom,rztcom,Ytcom,MIMO,DD,ISC}. It involves a 4D tensor CPD with three Vandermonde factor matrices.  To determine the uniqueness condition for our new formulation, we extend that derived in \cite{vand} to the decomposition of a fourth-order tensor with three Vandermonde factor matrices. This extension is presented in Lemma 1.

\begin{theorem} Considering a fourth-order tensor $\bc X = \llbracket \mf {A}_{(1)}, \mf A_{(2)},\mf A_{(3)}  ,\mf A_{(4)} \rrbracket \in \mb C^{I_1 \times I_2 \times I_3 \times I_4}$ with factor matrices $\mf {A}_{(1)} \in \mb C^{I_1 \times I_L}, \mf A_{(2)} \in \mb C^{I_2 \times I_L}, \mf A_{(3)} \in \mb C^{I_3 \times I_L} , \mf A_{(4)} \in \mb C^{I_4 \times I_L} $,  among which $\mf {A}_{(1)},\mf A_{(2)} $,$ \mf A_{(3)}$ are Vandermonde constrained with generators $\{z_{1,l}\}_{l=1}^{I_L}, \{z_{2,l}\}_{l=2}^{I_L}, \{z_{3,l}\}_{l=1}^{I_L}$.\footnote{{ The definition of Vandermonde generators is given in Eq.(9) of \cite{vand}.}} Under the constraints of  smooth parameters $(K_1,L_1,K_2,L_2,K_3,L_3)$ satisfying $K_1+L_1=I_1+1,K_2+L_2=I_2+1, K_3+L_3=I_3+1$, if 
	\begin{equation}
		\label{eq19}
		\begin{split}
			&z_{1,m} \ne z_{1,n}, \forall m\ne n\\
			&\mf {r}(\mf{A}_{(K_1-1,1)} \odot \mf{A}_{(K_2,2)} \odot \mf{A}_{(K_3,3)} )=I_L\\
			&\mf {r}(\mf{A}_{(L_1,1)} \odot \mf{A}_{(L_2,2)}\odot \mf{A}_{(L_3,3)} \odot \mf{A}_{(4)})=I_L, 
		\end{split}
	\end{equation}	
	then  the  CPD of $\bc X$ is unique. Specifically, $\mf{A}_{(K_1-1,1)}$ denotes the first $K_1-1$ rows of $\mf{A}_{(1)}$. Generically, condition (\ref{eq19}) equals to:
	\vspace{-2mm}
	\begin{equation}
		\label{eq20}
		\begin{split}
			&\min_{\{K_i+L_i=I_i+1\}_{i=1}^3} ((K_1-1)K_2K_3,L_1L_2L_3I_4) \ge I_L.
		\end{split}
	\end{equation}
\end{theorem}
\vspace{-1mm}

After comparing Lemma 1 to Corollary III.4 in \cite{vand}, it can be found that by incorporating an additional factor matrix with Vandermonde structure, the uniqueness condition can be significantly relaxed, thus allowing for the identification of hundreds of sub-paths. To ensure the uniqueness guaranteed by Lemma 1, we are to develop an algorithm explicitly utilizing the Vandermonde structures of the three factor matrices. 
To achieve this, we extend the algorithm in \cite{vand} and propose  the \textit{Vandermonde Structured Decomposition algorithm for Fourth-order tensors} (VSD-Fort). The key steps of this enhanced algorithm are outlined below. 

Consider $\bc X = \llbracket \mf {A}_{(1)}, \mf A_{(2)},\mf A_{(3)}  ,\mf A_{(4)} \rrbracket  \in \mb{C}^{I_1 \times I_2 \times I_3 \times I_4     }$   with three factor matrices $\{ \mf {A}_{(1)}, \mf A_{(2)},\mf A_{(3)} \}$ being Vandermonde structured,  choose pairs $(K_1, L_1), (K_2, L_2), (K_3, L_3)$ subject to $K_1+L_1=I_1+1$,$K_2+L_2=I_2+1$, $K_3+L_3=I_3+1$ firstly. Then hankerize $ \bc X$ and obtain ${\mf X}^{\text{hank}}$:
\begin{equation}
	\label{eq21}
	\begin{split}
		&{\mf X}^{\text{hank}}=( \mf{A}_{(K_1,1)} \odot  \mf{A}_{(K_2,2)} \odot  \mf{A}_{(K_3,3)}  )
		(\mf{A}_{(L_1,1)} \odot \\
		&\mf{A}_{(L_2,2)}  \odot  \mf{A}_{(L_3,3)} \odot \mf{A}_{(4)} )^{\T} \in \mathbb{C}^{K_1K_2K_3 \times L_1L_2L_3I_4 }.
	\end{split}
\end{equation}
Secondly, compute the SVD of ${\mf X}^{\text{hank}}$:
\begin{equation}
	\label{eq332}
	\text{SVD}({\mf X}^{\text{hank}})=\mf U \mf{\Sigma} \mf V^{\HH}=\mf U_l \mf{\Sigma}_l \mf V^{\HH}_l + \mf U_n \mf{\Sigma}_n \mf V^{\HH}_n,
\end{equation}
where $\mf U_l \mf{\Sigma}_l \mf V^{\HH}_l$ denotes signal subspace and $\mf U_n \mf{\Sigma}_n \mf V^{\HH}_n $ denotes noise subspace. Determine the potential rank of signal subspace ($R$)  from  $\mf{\Sigma}$ by enumerating significant  eigenvalues.
Then obtain sliced matrices $\mf U_1,\mf U_2$ 
\begin{equation}
	\label{eq482}
	\begin{split}
		&\mf U_1= \mf U(1:(K_1-1)K_2K_3, 1:R) ,\\
		&\mf U_2= \mf U(1+K_2K_3:K_1K_2K_3, 1:R),
	\end{split}
\end{equation}
and  compute the eigen decomposition of $\mf U_1^{\dagger}\mf U_2$, $\text{EVD}(\mf U_1^{\dagger}\mf U_2)=\mf M \mf Z \mf M^{-1}$.
The first Vandermonde generators set $\{z_{1,r}\}_{r=1}^{R}$ of $\mf {A}_{(1)}$ can thus be determined by normalizing diagonal entries of $\mf Z$ :
\begin{equation}
	\label{eq484}
	\begin{split}
		&\{z_{1,r}\}_{r=1}^{R}=\text{diag}(\mf Z), z_{1,r}=	\frac{z_{1,r}}{\left| z_{1,r}\right|}, r=1, ..., R.
	\end{split}	
\end{equation}
Then, reconstruct $\mf {\hat{A}}_{(1)}$ with $\{z_{1,r}\}_{r=1}^{R}$ and  compute  $(\mf{\hat{A}}_{(K_2,2)} \odot \mf{\hat{A}}_{(K_3,3)})$ through:
\begin{equation}
	\label{eq27}
	\begin{split}
		(\mf{\hat{A}}_{(K_2,2)} \odot \mf{\hat{A}}_{(K_3,3)})_{(:,r)}= &(\mf{\hat {a}}_{(K_1,r,1)}^{\HH} \otimes \mf{I}_{K_2K_3} ) \mf{U}\mf{m}_r
	\end{split}
\end{equation}
where $\mf{\hat {a}}_{(K_1,r,1)}$ denotes $\mf{\hat {A}}_{(1)}(1:K_1,r)$.
Then, determine  the second  Vandermonde generators set $\{z_{2,r}\}_{r=1}^{R}$ of $\mf{\hat{A}}_{(2)}$ :
\begin{equation}
	\label{eq486}
	\begin{split}
		&z_{2,r}=(\mf{\hat{A}}_{(K_2,2)} \odot \mf{\hat{A}}_{(K_3,3)}  )^{\dagger}_{(1:(K_2-1)K_3,r)}\\
		&(\mf{\hat{A}}_{(K_2,2)} \odot \mf{\hat{A}}_{(K_3,3)} )_{(K_3+1:K_2K_3,r )}.
	\end{split}
\end{equation}
Similarly, reconstruct $\mf{\hat{A}}_{(2)}$ with $\{z_{2,r}\}_{r=1}^R$. So far, we have obtained $\{\mf{\hat{A}}_{(1)}, \mf{\hat{A}}_{(2)}\}$, and $r$-th vector of $\mf{\hat{A}}_{(3)}$ can be derived: 
\begin{align}
	\label{eq1212}
	\mf{\hat{a}}_{(K_3,r,3)}=(\mf{\hat{a}}^{\HH}_{(K_2,r,2)}   \otimes \mf{I}_{K_3} ) &(\mf{\hat {a}}_{(K_1,r,1)}^{\HH} \otimes \mf{I}_{K_2K_3} ) \mf{U}\mf{m}_r
\end{align}
Given estimated  $\mf{\hat{A}}_{(K_3,3)}$, the third   Vandermonde generators set $\{z_{3,r}\}_{r=1}^{R}$ can be extracted for recorrection. Given three estimated factor matrices,  $\mf{A}_{(4)}$ has a closed-form solution:
\begin{equation}
	\label{eq28}
	\begin{split}
		\mf{\hat{A}}_{(4)}=&\mf{X}_{(4)}( \mf {\hat{A}} _{(3)} \odot \mf {\hat{A}}_{(2)} \odot \mf {\hat{A}}_{(1)} )\\
		&(\mf {\hat{A}}_{(3)}^{\T}\mf {\hat{A}}_{(3)} \ast \mf {\hat{A}}_{(2)}^{\T}\mf {\hat{A}}_{(2)} \ast \mf {\hat{A}}_{(1)}^{\T}\mf {\hat{A}}_{(1)}  )^{\dagger},
	\end{split} 
\end{equation}
where $\mf{X}_{(4)}$ represents mode-4 unfolding of $\hat{\bc {X}}$. 
\begin{algorithm}[!t]
	\caption{ : VSD-Fort for Estimating Uplink Channels} 
	\hspace*{0.02in} {\bf Input:} 
	$\bc {\hat{H}}_{\text{comb}} \in \mb{C}^{N_{\text{col}} \times N_{\text{row}} \times N_{\text{sc,eff}} \times N_{\text{pol}}        }   $.
	\label{a1}
	\begin{algorithmic}[1]
		\State Compute the factor matrices $\mf{\hat{A}}(\phi),\mf{\hat{A}}(\theta),\mf{\hat{D}}_{\text{comb}}, \mf{\hat{P}}$ via (\ref{eq21})$\sim$(\ref{eq28});
		\State Compute $\{\hat{\tau}_l^u\}_{l=1}^{L}$ through the logarithm of Vandermonde generators of $\mf{\hat{D}}_{\text{comb}}$ and then reconstruct  $\mf{\hat{D}}$ via (\ref{model}).
		
		\State Reconstruct $\bc{\hat{H}}$ via (\ref{eq8}).
	\end{algorithmic}
	\hspace*{0.02in} {\bf Output:} 
	$\bc {\hat{H}} \in \mb{C}^{N_{\text{col}} \times N_{\text{row}} \times K \times N_{\text{pol}} }$
\end{algorithm}
So far, we have presented our proposed VSD-fort algorithm. In this context, we treat the received signal matrix $\bc X$ as  $\bc{\hat{H}}_{\text{comb}}^u$, and the matrices to be estimated  $\{\mf{\hat{A}}(\phi),\mf{\hat{D}}_{\text{comb}},\mf{\hat{A}}(\theta), \mf{\hat{P}}\}$ as $\{\mf {A}_{(1)}, \mf A_{(2)},\mf A_{(3)}  ,\mf A_{(4)}\}$. We can estimate the complete channel  based on the model in (\ref{model}) after reconstructing   $\mf{\hat{D}}$  following  (\ref{eq8}) with the estimated Vandermonde generators, as summarized in \textbf{Algorithm \ref{a1}}.
\vspace{-3mm}
	\begin{table}[h]
	\begin{center}
		\caption{Simulation Parameters }
		\label{t1}
		\begin{tabular}{|c|c|} 
			\hline
			\rowcolor{gray!30}	
			\parbox[c][0.4cm][c]{3cm} {\centering \textbf{Parameters} } &\textbf{Value} \\
			\hline	
			\parbox[c][0.3cm][c]{3cm} {\centering Modulation type} &	{\scriptsize 64QAM} \\
			\hline
			\parbox[c][0.3cm][c]{3cm} {\centering BS Antenna} &{\scriptsize  4 $\times $16 UPA, $\pm 45^{\circ}$ cross-polarized} \\
			\hline
			\parbox[c][0.3cm][c]{3cm} {\centering UE Antenna} & {\scriptsize single isotropic $\times$ 24UEs} \\
			\hline
			\parbox[c][0.3cm][c]{3cm} {\centering	Carrier Frequency} & {\scriptsize 4.9GHz} \\
			\hline
			\parbox[c][0.3cm][c]{3cm}{\centering	Subcarrier space} & {\scriptsize 30KHz} \\
			\hline
			\parbox[c][0.3cm][c]{3cm} {\centering	Bandwidth} & {\scriptsize 20MHz} \\
			\hline
			\parbox[c][0.3cm][c]{3cm}{\centering {\tiny Subcarriers allocated for transmission}} &{\scriptsize  $K$: 384} \\
			\hline 
			\parbox[c][0.3cm][c]{3cm} {\centering {\tiny Effective subcarriers allocated for  comb}} & {\scriptsize $N_{\text{sc,eff}}$: 32} \\
			\hline
			\parbox[c][0.3cm][c]{3cm} {\centering {\tiny Antenna spatial interval} {\tiny $  d_{\text{col}}$, $d_{\text{row}}$}   } &{\scriptsize  3cm, 9cm}  \\
			\hline
		\end{tabular}
	\end{center}
\end{table}
\vspace{-8mm} 
\section{Simulation Results and Discussions}
In this section, we present numerical results to evaluate the performance of the proposed algorithm. Detailed system parameters are provided in Table \ref{t1}, and our simulations strictly adhere to the  3GPP protocols \cite{3GPP38901,3GPP38214,3GPP38211}. According to the settings specified in Table I, our proposed algorithm can identify a maximum of 420 sub-paths, as derived by Lemma 1.

\begin{figure}
	\centering
	\includegraphics[width=\linewidth]{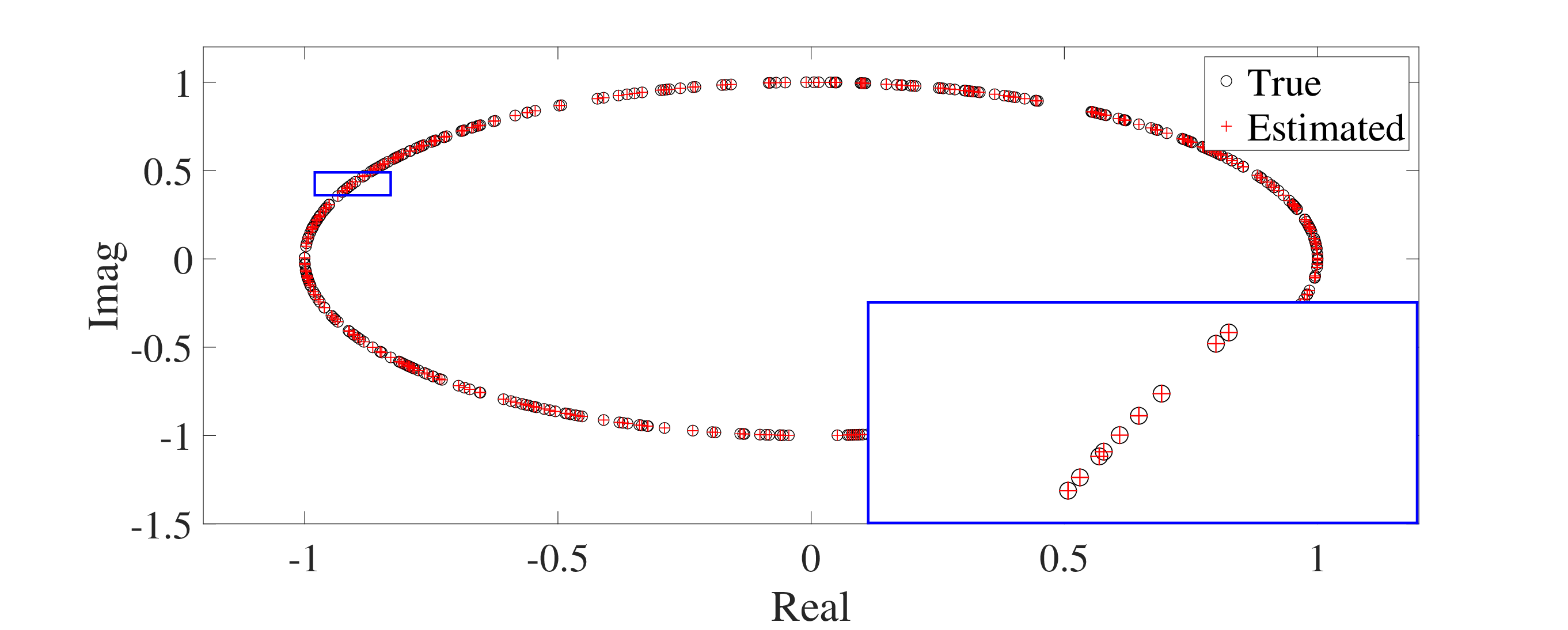}
	\caption{{ Illustration of estimations of the first set of Vandermonde generators of a high rank (420) CP tensor $\bc H \in \mb{C}^{16 \times 32 \times 4 \times 2}$ using our proposed method. } }
	\label{fig:vand}
\end{figure}

To validate the algorithm's ability to resolve channels with a large number of sub-paths, we simulate a channel  containing 420 sub-paths and utilize the VSD-Fort to estimate the first set of  Vandermonde generators  characterizing the channel. The results displayed in Fig. \ref{fig:vand} show that  estimated Vandermonde generators perfectly match all the ground-truth values. This validates  our method being able to  accurately estimate Vandermonde generators even in scenarios involving hundreds of sub-paths.
\begin{figure}
	\centering
	\includegraphics[width=\linewidth]{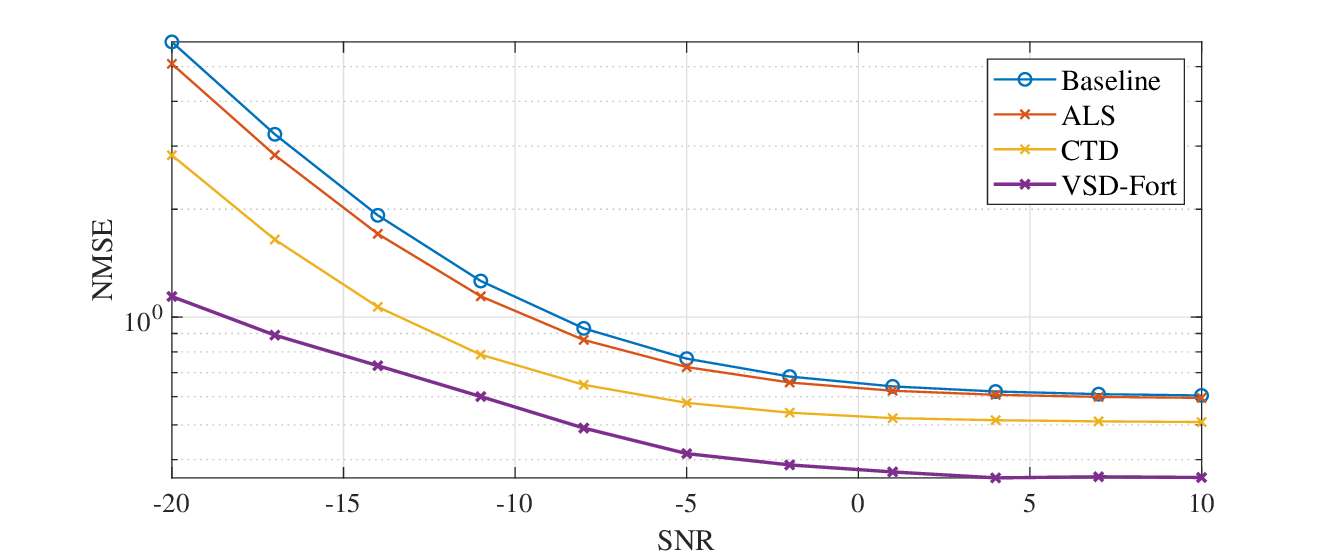}
	\caption{The channel estimation performance of different algorithms versus different SNRs.}
	\label{fig:test2}
\end{figure}
Then we evaluate the performance of channel estimation using the normalized mean squared error (NMSE) as the performance metric: $\text{NMSE}=\left\| \bc H-\bc{\hat{H}} \right \|_{\text{F}}^2 \big/ \left\| \bc H \right \|_{\text{F}}^2$, where $\bc H$ denotes the ground-truth channel and $\bc{\hat{H}}$ denotes the estimated channel. We consider a typical  MU-MIMO uplink transmission scenario with two cells, each consisting of 24 users. The BS in the targeted cell experiences strong interference from the adjacent cell (the power of target signals equal to that of targeted ones). { The parameters of  channels in the target cell follow the specifications of the  3GPP TR38.901 protocol (more details can be found in Table 7.7.1-3 of \cite{3GPP38901})  and each channel possesses 360 sub-paths. We add  random phase biases to the above parameters to create interference channels of  adjacent cell to the base station in the target cell.}

Fig. \ref{fig:test2} presents the averaging NMSEs of the baseline\cite{3GPP38214}, the ALS-based method\cite{ALS}, the CTD method\cite{DD} and our proposed VSD-Fort algorithm over 1000 trials, with varying signal-to-noise ratios (SNRs). The results  indicate that our proposed VSD-Fort algorithm outperforms the others in channel estimation,  { thus further contributing to  interference cancellation and  reduced symbol detection error\cite{detection}.}

\section{Conclusions}
\label{sec5}
In this letter,  we introduce a structured high-rank tensor approach for sub-6G uplink channel estimation in the  MU-MIMO communications. By exploring the physical structure of   channel, we represent it as a fourth-order CP tensor with three Vandermonde factors. Upon its stronger identifiability, we propose the VSD-Fort  to acquire parameters of the channel and subsequently obtain the entire channel estimations. Numerical experiments  demonstrates that our proposed algorithm  achieves superior performance with robustness against noise and interference. Additionally, our method exhibits plug-in compatibility with the widely-used baseline approach, rendering it highly potential for practical applications.

\end{document}